\documentclass[aps,prl,reprint,superscriptaddress,longbibliography,showpacs,showkeys]{revtex4-1}

\usepackage{amsmath,bm,bbm,amssymb}
\usepackage[squaren]{SIunits}
\usepackage[usenames,dvipsnames]{color}
\usepackage[pdftex]{graphicx}
\usepackage{ulem}
\usepackage{hyperref}
\usepackage[english]{babel}

\usepackage[resetlabels]{multibib}
\newcites{SI}{nada}

\newcommand{\omegaL}{\omega_{\rm L}}

\newcommand{\mcite}[1]{~\cite{#1}}
\newcommand{\pcite}[1]{~\cite{#1}.}
\newcommand{\ccite}[1]{~\cite{#1},}

\newcommand{\mciteSI}[1]{~\citeSI{#1}}
\newcommand{\pciteSI}[1]{~\citeSI{#1}.}
\newcommand{\cciteSI}[1]{~\citeSI{#1},}

\newcommand{\ave}[1]{\ensuremath{\langle#1\rangle}}
\newcommand{\norm}[1]{\ensuremath{\left|#1\right|}}

\newcommand{\cov}{\mathrm{cov}}

\newcommand{\var}{\mathrm{var}}

\newcommand{\myhat}[1]{#1}

\newcommand{\myF}{\myhat{F}}
\newcommand{\myf}{\myhat{f}}
\newcommand{\myS}{\myhat{S}}

\newcommand{\myx}{x}
\newcommand{\myy}{y}
\newcommand{\myz}{z}

\newcommand{\myyz}{{y,z}}

\newcommand{\Fx}{\myF_\myx}
\newcommand{\Fy}{\myF_\myy}
\newcommand{\Fz}{\myF_\myz}
\newcommand{\Fyz}{\myF_\myyz}

\newcommand{\fz}{\myf_\myz}
\newcommand{\Fa}{\myF_{\alpha}}
\newcommand{\fa}{\myF_{\alpha}}

\newcommand{\Sx}{\myS_\myx}
\newcommand{\Sy}{\myS_\myy}
\newcommand{\Sz}{\myS_\myz}
\newcommand{\Sk}{\myS_k}

\newcommand{\supin}{^{({\rm in})}}
\newcommand{\supout}{^{({\rm out})}}

\newcommand{\myvphi}{\myhat{\varphi}}

\newcommand{\vph}{\ensuremath{\myvphi}}
\newcommand{\vPh}{\Phi}

\newcommand{\myomit}[1]{}

\newcommand{\NA}{N_{\rm A}}
\newcommand{\NL}{N_{\rm L}}

\newcommand{\NLH}{N_{\rm L}^{\rm (H)}}

\newcommand{\Npulse}{N_{\rm p}}

\newcommand{\rb}{$^{87}\mathrm{Rb}\ $}
\newcommand{\gs}{\ensuremath{5{\rm S}_{1/2}}}
\newcommand{\es}{\ensuremath{5{\rm P}_{3/2}}}

\newcommand{\F}{\mathbf{F}}

\newcommand{\tpoint}{t_k}
\newcommand{\tdelta}{\Delta t}
\newcommand{\tmeas}{t_{\rm e}}
\newcommand{\trel}{t_{\rm r}}

\newcommand{\offset}{\vph_0}

\newcommand{\DeltaMeas}{\Delta_{\rm m}}
\newcommand{\DeltaDist}{\Delta_{\rm d}}

\begin{document}

\title{Simultaneous tracking of spin angle and amplitude beyond classical limits}

\newcommand{\ICFOAddress}{ICFO-Institut de Ciencies Fotoniques, The Barcelona Institute of Science and Technology, 08860 Castelldefels (Barcelona), Spain}
\newcommand{\ICREAAddress}{ICREA -- Instituci\'{o} Catalana de Re{c}erca i Estudis Avan\c{c}ats, 08015 Barcelona, Spain}

\author{Giorgio~Colangelo}
\email[]{giorgio.colangelo@icfo.eu}
\affiliation{\ICFOAddress}

\author{Ferran~Martin Ciurana}
\affiliation{\ICFOAddress}

\author{Lorena~C. Bianchet}
\affiliation{\ICFOAddress}

\author{Robert~J. Sewell}
\affiliation{\ICFOAddress}

\author{Morgan~W.~Mitchell}
\affiliation{\ICFOAddress}
\affiliation{\ICREAAddress}

\begin{abstract}
We show how simultaneous, back-action evading tracking of non-commuting observables can be achieved in a widely-used sensing technology, atomic interferometry. Using high-dynamic-range dynamically-decoupled  quantum non-demolition (QND) measurements on a precessing atomic spin ensemble, we track the collective spin angle and amplitude with negligible effects from back action, giving steady-state tracking sensitivity $\unit{2.9}{dB}$  beyond the standard quantum limit and $\unit{7.0}{dB}$ beyond Poisson statistics.
\end{abstract}

\date{\today}
\maketitle

Continuous monitoring or \textit{tracking} of a quantum system is essential to high-sensitivity measurement of time-varying quantities\mcite{TsangPRL2011} from biomagnetic fields\cite{KominisN2003OLD} to gravitational-wave strain\pcite{LIGOPRL2016a}
Naive tracking strategies have limited sensitivity due to quantum back-action, in which measurement of one observable disturbs other, non-commuting observables\pcite{BraginskyS1980} Quantum-aware strategies have shown back-action evasion, foregoing knowledge of one observable to precisely measure another\pcite{GrangierN1998,RuskovPRB2005,SewellNP2013,VasilakisNatPhys2015}
Recent proposals suggest that back-action can be evaded even when tracking multiple, non-commuting observables, by employing negative-mass oscillators \mcite{TsangPRX2012,PolzikADP2015,MollerArxiv2016} or zero-area Sagnac interferometers\pcite{ChenPRD2003,GrafCQG2014} 
Here we show how simultaneous, back-action evading tracking of non-commuting observables can be achieved in a widely-used sensing technology, atomic interferometry. Using high-dynamic-range\ccite{MartinOL2016} dynamically-decoupled\mcite{KoschorreckPRL2010b} quantum non-demolition (QND) measurements\ccite{SewellNP2013,VasilakisNatPhys2015} on a precessing atomic spin ensemble, we track the collective spin angle and amplitude with negligible effects from back action, giving steady-state tracking sensitivity $\unit{2.9}{dB}$  beyond the standard quantum limit and $\unit{7.0}{dB}$ beyond Poisson statistics\pcite{BeguinPRL2014}
The technique greatly extends the quantum limits for atomic sensors that track frequency\ccite{LudlowRMP2015} acceleration, rotation and gravity\ccite{CroninRMP2009} magnetic fields\ccite{BudkerNP2007} and physics beyond the standard model\pcite{BudkerPRX2014}

In the context of gravitational-wave searches, it was noted\mcite{BraginskyS1980} that the uncertainty principle constrains not only our knowledge of quantum systems, but also of seemingly non-quantum observables such as time and distance. Because our instruments to measure these are necessarily quantum systems, they are subject to measurement back-action. 
In harmonic oscillators and optical modes, continuous measurement of one quadrature $X$ disturbs the other quadrature $P$, which through the dynamics of oscillation returns the disturbance to the measured variable. 
The effects of back-action can be evaded, however, if $X$ is measured stroboscopically, at the same phase each cycle. 
The disturbance to $P$ then never enters the measurement record, but also no information is gained about $P$. Varieties of this single-variable ``back-action evading'' or QND measurement have been implemented with photonic\ccite{GrangierN1998} mechanical\ccite{RuskovPRB2005} and atomic systems\pcite{SewellNP2013,VasilakisNatPhys2015} 

Recently, the possibility of evading measurement back-action for \textit{both} variables of an oscillator has been suggested.
This task, which might at first seem impossible, is attractive because it would allow back-action-unlimited detection of both amplitude- and phase-perturbing effects, and requires no \textit{a priori} knowledge of the oscillator phase. 
Existing proposals involve matched systems: when an ordinary oscillator is matched to a negative-mass counterpart, a subsystem becomes immune to the uncertainty principle but remains sensitive to external forces\pcite{OzawaPLA2004,TsangPRX2012,PolzikADP2015} 
In this way, the zero-area Sagnac interferometer is predicted to evade back-action in sensing gravitational waves\pcite{ChenPRD2003,GrafCQG2014} 

Here we show that back-action evasion can be achieved when tracking non-commuting observables in atomic sensors. 
Using quasi-continuous quantum non-demolition measurements, we track the two oscillating observables of a single macroscopic spin oscillator. 
The measurement evades all but a negligible back-action contribution, to obtain a record of both the amplitude and angle of the oscillator beyond their respective classical limits. 
This demonstrates for atomic interferometers a sensing modality unavailable to mechanical and optical oscillators and compatible with the most advanced atomic sensing strategies.

\begin{figure*}
\includegraphics[width=\textwidth]{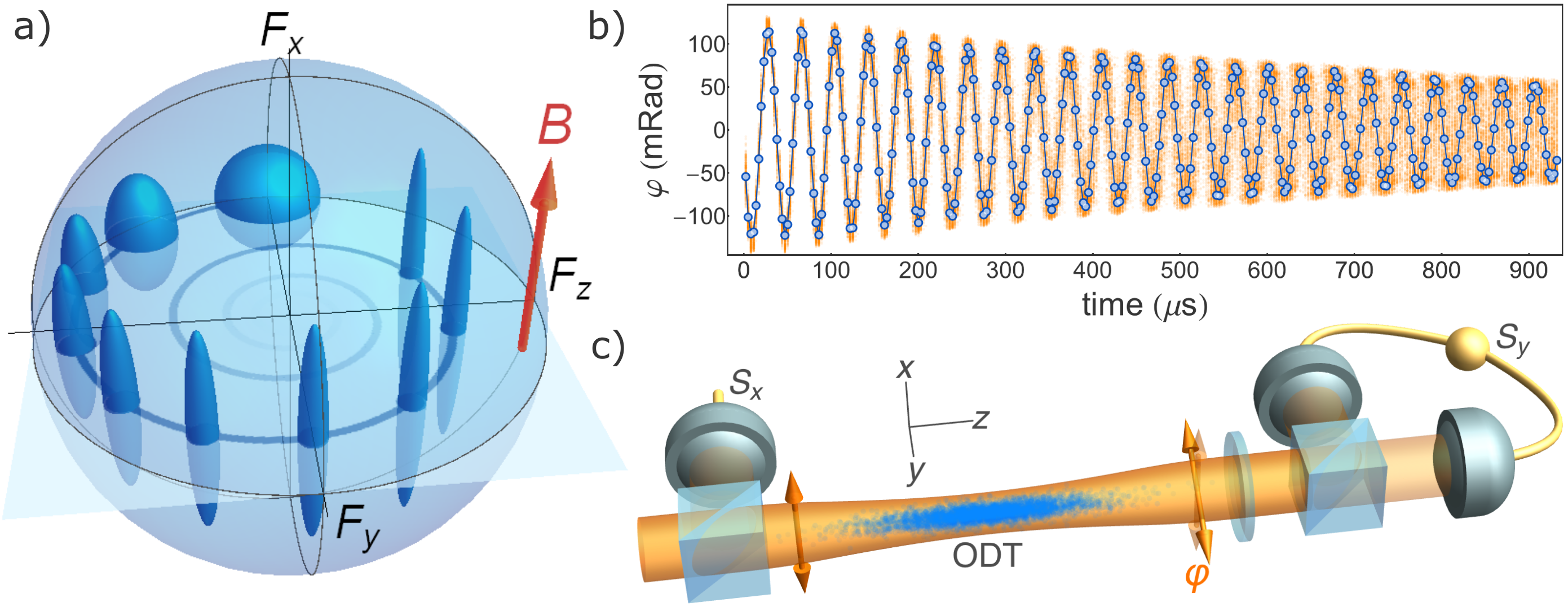}
\caption{
Simultaneous tracking of non-commuting observables. 
a) Bloch-sphere representation of the atomic state evolution. 
Ellipsoids show uncertainty volumes (not to scale) as the state evolves anti-clockwise from an initial, ${\Fy}$-polarized state with isotropic uncertainty. An $x$-oriented magnetic field ${\bf B}$ drives a coherent spin precession in the $\Fy$--$\Fz$ plane. 
Quasi-continuous measurement of $\Fz$ produces a reduction in $\Fz$ and $\Fy$ variances, with a corresponding increase in $\var(\Fx)$.
b) Observed Faraday rotation angle $\vph \propto \Fz$ versus time. Each circle shows the 
rotation angle from one V-polarized pulse. 
A magnetic field of $\unit{37.6}{\milli G}$ produces the observed oscillation, while dephasing due to residual magnetic gradients and off-resonant scattering of probe photons cause the decay of coherence.
Blue circles show a single, representative trace, overlaid on $450$ repetitions of the experiment shown as orange dots. 
The time zero corresponds to the first probe pulse; the end of optical pumping is \unit{58}{\micro\second} earlier. 
c) Experimental geometry: $1.88\times 10^6$ cold $^{87}$Rb atoms are confined in a weakly-focused single beam optical dipole trap (ODT). 
Transverse optical pumping is used to produce $\Fy$ polarisation. 
On-axis, \unit{0.6}{\micro \second} pulses with mean photon number $2.74\times 10^6$ experience Faraday rotation by an angle $\vph \propto \Fz$. 
A polarimeter consisting of waveplates, a polarising beamsplitter, high-quantum-efficiency photodiodes, and charge-sensitive amplifiers measures the output Stokes component $\Sy$. 
A reference detector before the atoms measures input Stokes component $S_0 = |\Sx|$. 
The rotation angle is computed as $\varphi = \arcsin(\Sy/\Sx)$. 
\label{fig:pqs}}
\end{figure*}

Atomic interferometers employ atomic ensembles that behave as a large spin ${\bf F}$ governed by an angular momentum algebra. 
As ${\bf F}$ precesses about any given axis, two spin components oscillate harmonically while the third is constant. 
Precessing about the $x$ axis, the oscillating components obey the Robertson uncertainty relation\mcite{RobertsonPR1929} 
\begin{equation}
\label{eq:Robertson}
\delta \Fy \delta \Fz \ge \frac{1}{2} | \langle [\Fy, \Fz] \rangle | = \frac{1}{2} | \langle \Fx \rangle |
\end{equation}
(we take $\hbar = 1$ throughout). 
For the best signal, polarization in the $\Fy$--$\Fz$ plane should be maximal, in which case $| \langle \Fx \rangle |$ vanishes. 
Because $\langle \Fx \rangle$ is a constant of the motion, this condition holds for all time and Eq.~(\ref{eq:Robertson}) sets no limit on the area in the $\Fy$--$\Fz$ phase space.
When the Robertson relation is thus evaded, arithmetic uncertainty relations\mcite{DammeierNJP2015} limit the uncertainty to $\var (\Fy) + \var (\Fz) \sim \NA^{2/3}$, far below $\sim\NA^{}$, the standard quantum limit (SQL)\pcite{HePRA2011} 
 $\NA$ is typically $\sim 10^6$ in cold atom systems and $\sim 10^{12}$ in atomic vapors, so this $\NA^{1/3}$ advantage extends the quantum limits by orders of magnitude.

Setting $|\langle \Fx \rangle|=0$ evades one obstacle to tracking, that presented by Eq.~(\ref{eq:Robertson}). 
We also must show that a non-destructive measurement of $\Fz, \Fy$ can be engineered to avoid back-action effects. 
In the Supplementary Information we show this for an ideal 
Faraday rotation measurement.
The state evolution is illustrated in Fig.~\ref{fig:pqs}~a) and summarized here: 
$\Fz$ is coupled to an optical ``meter'' variable $\Sz$ via the QND interaction $H_{\rm eff} = g \Fz \Sz$, where $g$ is a coupling constant. 
The interaction with $\NL$ photons imprints a signal proportional to $\Fz$ on the meter, which when measured reduces $\var(\Fz)$ by an amount $\DeltaMeas \sim g^2  \NL \var^2(\Fz)$. 
This same interaction rotates ${\bf F}$ about $\Fz$ by a random angle $\theta \equiv g \Sz$, increasing $\var(\Fy)$ by an amount $\DeltaDist \sim g^2  \NL \var(\Fx)$, and also increasing $\var(\Fx)$.
Precessing and under continuous measurement, $\Fz$ and $\Fy$ alternate roles as the measured and disturbed variable, and each experiences both effects. 
For $\overline{\mathrm{N}}_{\mathrm{L}}=1/(g^2 \NA)$, the measurement benefit $\DeltaMeas \sim \NA$ is of order the initial uncertainty, while the back-action $\DeltaDist \sim 1$ is negligible.
Probing with this $\NL$ induces a negligible loss of coherence, so that the
sensitivity to both angular and radial perturbations improves.

\begin{figure*}
\includegraphics[width=\textwidth]{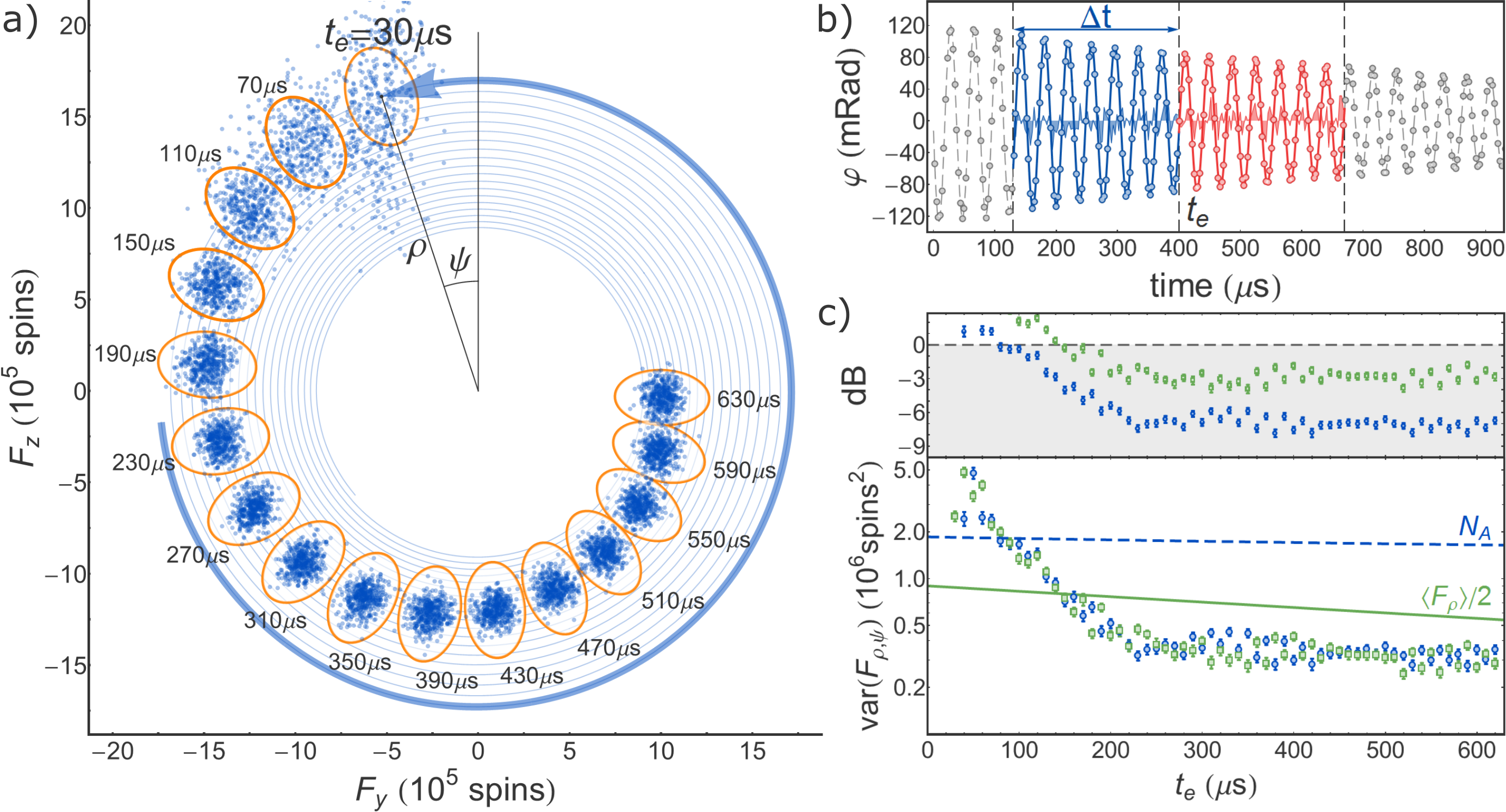}
\caption{
Experimental results. 
a) Measured trajectories in the $\Fy$--$\Fz$ phase space at different estimation times $\tmeas$. 
For each of the 450 traces shown in Fig.~\ref{fig:pqs}~b), the function of Eq.~(\ref{eq:FIDForm}) is {fit to the data} to find predictive and confirming estimates $\F_1$, $\F_2$, respectively, for $(\Fy,\Fz)$ at time $\tmeas$.
Fits for $\F_1$ and $\F_2$ use disjoint sets of data covering the ranges
$\tmeas-\tdelta \le \tpoint < \tmeas$ and $\tmeas < \tpoint \le \tmeas+\tdelta$
, respectively. 
A single fit is a tightly-wound spiral shown as a thin blue line 
and the thick arrow shows the trajectory from $t=0$ to $t=\tmeas=\unit{30}{\micro\second}$.
For clarity, we show results for $\tmeas$ values spaced by \unit{40}{\micro \second}, slightly more than one Larmor oscillation. 
Each point shows $\langle \F_1 \rangle + 100 \mathcal{F}$, where $\langle \F_1 \rangle$ is the mean over the 450 repetitions, and ${\bf \mathcal{F}} \equiv \F_2 - \Gamma_{2,1} \Gamma_{1}^{-1} \F_1$ is the error of the best linear prediction (see SI). 
The factor 100 provides magnification for visualization purposes. 
Orange ellipses, with radial and azimuthal radii of $2\sigma$, where $\sigma=100\sqrt{\mathrm{CL}}$, show the relevant classical limits: Poisson (radial, $\mathrm{CL}=\NA$) and SQL (azimuthal, $\mathrm{CL}=\langle F_{\rho} \rangle/2$).
b) Fits to estimate $(\Fy,\Fz)$ for $\tmeas = \unit{400}{\micro \second}$ and a measurement time $\Delta t=\unit{270}{\micro \second}$. 
Blue (red) shows fits based on prior (posterior) data. 
Shaded regions show fit residuals $\times 10$. 
c) Evolution of tracking precision for different $\tmeas$. 
Blue circles and green squares show radial and azimuthal components of $\Gamma_{\F_2\mid \F_1}$. 
Error bars show $\pm 1\sigma$ standard error. 
Dashed blue and solid green curves show Poisson and SQL variances. 
These decrease during probing due to loss of coherence and loss of atoms.
No readout noise has been subtracted.
\label{fig:PLPQS}}
\end{figure*}

Realizing this in-principle advantage requires control of measurement dynamics\mcite{SmithPRL2004} and incoherent effects\ccite{KoschorreckPRL2010a} as well as low-noise non-destructive detection with high-dynamic-range\pcite{MartinOL2016} 
We use an ensemble of $\NA=1.88\times 10^6$ cold $^{87}$Rb atoms held in an optical dipole trap. 
The atoms are initially prepared in the $\Fy$-polarized state by optical pumping and, due to an applied B-field in the $x$ direction, precess coherently in the $\Fy$--$\Fz$ plane with Larmor period $T_L \approx \unit{38}{\micro\second}$.
The ``meter'' variable is the polarisation of \unit{\sim 1}{\micro \second}, off-resonance optical pulses, which experience Faraday rotation by an angle $\varphi = g \Fz$ on the Poincare sphere as they propagate through the atomic cloud. 
We probe the atoms with V-polarized optical pulses, interspersed with H-polarized compensation pulses
to dynamically decouple the spin alignment\ccite{KoschorreckPRL2010b,SewellNatPhot2013} i.e., to produce the effective hamiltonian $H_{\rm eff}=g \Fz \Sz$ without tensor light shifts. 
We use high dynamic-range, shot-noise-limited optoelectronics\mcite{MartinOL2016} and nonlinear signal reconstruction to achieve sub-projection-noise readout sensitivity for rotation signals up to $\varphi \approx \unit{100}{\milli\radian}$. 
See Supplementary Information.

A representative sequence of measured Faraday rotation angles  $\vph(\tpoint$) for QND measurements spread over \unit{1}{\milli \second} is shown in Fig.~\ref{fig:pqs}~b), and is well described by a free induction decay model that we use to estimate $\Fz$ and $\Fy$ at a time $\tmeas$
\begin{equation}
\label{eq:FIDForm}
\vph(t)=g \left[\Fz(\tmeas) \cos \omegaL \trel -\Fy(\tmeas) \sin \omegaL \trel \right] e^{-\trel/T_2} + \offset
\end{equation}
where $\trel \equiv t- \tmeas$.
The coupling constant $g$ is found by an independent calibration, while the Larmor frequency $\omegaL$, the coherence time $T_2$, and the offset $\offset$ are found by fitting to the measured $\vph(\tpoint)$ over the full range of $t$. 

With these parameters fixed, we then use Eq.~(\ref{eq:FIDForm}) to obtain a \textit{predictive} estimate $\F_1 = (\Fy^{(1)}, \Fz^{(1)})$ at time $\tmeas$ using measurements $\{ \varphi(\tpoint) \}_{\tmeas-\tdelta \le \tpoint < \tmeas}$; and  to obtain a \textit{confirming} estimate $\F_2 = (\Fy^{(2)}, \Fz^{(2)})$ using $\{\varphi(\tpoint)\}_{\tmeas < \tpoint \le \tmeas + \tdelta}$.
Because the classical parameters $g$, $\omegaL$, $T_2$ and $\offset$, are fixed beforehand, these are two linear, least-squares estimates of the vector $\F$ obtained from disjoint data sets. 
Estimating $\F$ for several values of $\tmeas$ gives a predictive trajectory and a confirming one. 
We gather statistics over 450 repetitions of the experiment.
Empirically, we find $\tdelta = \unit{270}{\micro\second}$ minimizes the total variance ${\rm Tr}(\Gamma_{\F_2\mid \F_1})$ (see Supplementary Information), reflecting a trade-off of photon shot noise versus scattering-induced decoherence and magnetic-field technical noise.

Fig.~\ref{fig:PLPQS}~a) shows the resulting mean predictive trajectory $\langle \F_1 \rangle$, which spirals slowly toward the origin due to magnetic-gradient-induced dephasing, and the discrepancy between the trajectories, $\F_2 - \F_1$, which rapidly decreases due to the measurement effect, reaching a steady state after about \unit{250}{\micro \second} of probing.

To quantify the measurement uncertainty, we compute the vector conditional covariance $\Gamma_{\F_2\mid \F_1}=\Gamma_{\F_2}-\Gamma_{\F_{2} \F_1} \Gamma_{\F_1}^{-1}\Gamma_{\F_1 \F_2}$ where $\Gamma_{\bf v}$ indicates the covariance matrix for vector ${\bf v}$, and $\Gamma_{\bf uv}$ indicates the cross-covariance matrix for vectors ${\bf u}$ and ${\bf v}$. 
Defining the polar coordinate system $(\Fy,\Fz) = \rho (-\sin \psi,\cos \psi)$, we identify the radial and azimuthal variances, $\var(F_\rho) \equiv \hat{\rho}^T \Gamma_{\F_2\mid \F_1} \hat{\rho}$ and $\var(F_\psi) \equiv \hat{\psi}^T \Gamma_{\F_2\mid \F_1} \hat{\psi}$, respectively, where $\hat{\rho} \equiv (-\sin \psi, \cos \psi)^T$ and $\hat{\psi} \equiv (-\cos \psi,-\sin \psi)^T$ are radial and azimuthal unit vectors. 

As shown in Fig.~\ref{fig:PLPQS}~c), $\var(F_\psi)$ drops below the SQL of $\langle F_{\rho} \rangle/2$
after $\approx \unit{150}{\micro \second}$ of probing, and remains below it to the limit of the experiment. No read out noise has been subtracted. 
Considering the steady-state region $\tmeas\geq \unit{270}{\micro \second}$, $\var(F_\psi)$ is on average \unit{2.9}{dB} below the SQL, and $\var(F_\rho)$  is on average \unit{7.0}{dB} below the Poissonian variance $\NA$, {to give a precision surpassing classical limits in both dynamical variables}. For any given value of $\tmeas$, $\var(F_\rho)$  and $\var(F_\psi)$ have standard errors of $\approx \unit{0.3}{dB}$, implying high statistical significance even without combining results for different $\tmeas$. 

We have shown how measurement back-action can be made negligible in high-sensitivity atom interferometry, allowing continuous tracking of the full dynamics of non-commuting spin observables beyond classical limits. 
The method is very close to practical application in the highest-performance atomic sensors: 
Tracking of atomic spin precession by non-destructive optical measurement is already used in the highest-sensitivity magnetic field measurements\mcite{KominisN2003OLD}, and in some optical lattice clocks\pcite{LodewyckPRA2009}
Moreover, multi-pass\mcite{ShengPRL2013} and cavity build-up\mcite{HostenN2016} methods are compatible with these techniques and greatly reduce scattering-induced decoherence, the limiting factor in our experiment. 
Together, these advances enable tracking far beyond the standard quantum limit with atomic sensors. 

\section*{Acknowledgements}
We thank G. Vitagliano, M. D. Reid, P. D. Drummond, G. T\'oth, N. Behbood, M. Napolitano, S. Palacios, X. Menino and the ICFO mechanical workshop, J.-C. Cifuentes and the ICFO electronic workshop, D. T. Campbell and M. M. Fria. Work supported by MINECO/FEDER, MINECO projects MAQRO (Ref. FIS2015-68039-P), XPLICA (FIS2014-62181-EXP) and Severo Ochoa grant SEV-2015-0522, Catalan 2014-SGR-1295, by the European Union Project QUIC (grant agreement 641122), European Research Council project AQUMET (grant agreement 280169) and ERIDIAN (grant agreement 713682), and by Fundaci\'{o} Privada CELLEX. LCB was supported by the International Fellowship Programme `La Caixa' - Severo Ochoa, awarded by the `La Caixa' Foundation.

\clearpage

\section*{Supplementary Information}

\subsection{Faraday rotation probing of atomic spins}
\label{sec:FarRot}

The effective atom-light interaction is given by the hamiltonian
\begin{equation}
H_{\rm eff} = g \Sz \Fz
\label{qnd}
\end{equation}
which describes a quantum non-demolition measurement of the collective atomic spin $\Fz$, where the operators $\Fa \equiv \sum_{i} \fa^{(i)}$ (with $\alpha=x,y,z$) describe the collective atomic spin, with $\fa^{(i)}$ the spin orientation of individual atom spins. 
The optical polarization of the probe pulses is described by the Stokes operators $\Sk = \frac{1}{2}(a_L^\dagger, a_R^\dagger) \sigma_k (a_L, a_R)^T$, with Pauli matrices $\sigma_k$.
The coupling constant $g$ depends on the detuning from the resonance, the atomic structure and the geometry of the atomic ensemble and probe beam and is independently measured \citeSI{KubasikPRA2009,KoschorreckPRL2010a,Deutsch2010OC,KuzmichPRL2000,AppelPNAS2009} .

An input $\Sx$-polarized optical pulse interacting with the atoms experiences a rotation by an angle $\varphi = g \Fz$ because of the interaction given by eq. \eqref{qnd}.
The transformation produced by the measurement on $\Sy$ is
\begin{eqnarray}
\Sy' & = & \Sy \cos \varphi + \Sx \sin \varphi 
\end{eqnarray}
In our experiment we measure $\Sx$ at the input by picking off a fraction of the optical pulse and sending it to a reference detector, and $\Sy'$ using a fast home-built balanced polarimeter \citeSI{MartinOL2016}. 
Both signals are recorded on a digital oscilloscope.

From the record of $\Sx$ and $\Sy'$, we calculate $\hat{\varphi}$, the estimator for $\varphi$:
\begin{eqnarray}
\hat{\varphi}&=&\arcsin \left( \frac{\Sy'}{\Sx} \right)
\nonumber \\ 
& = & \arcsin \left( \sin \varphi + \frac{\Sy}{\Sx} \sqrt{1-\sin^2 \varphi} \right)
\nonumber \\ 
& = & \varphi + \frac{\Sy}{\Sx} 
+ \frac{1}{2} \left(\frac{\Sy}{\Sx}\right)^2 \tan \varphi 
+ O\left(\frac{\Sy}{\Sx}\right)^3.
\label{eq:arcsin}
\end{eqnarray}
We note that due to shot noise $\Sy/\Sx$ is normally distributed with zero mean and variance $1/(2 \Sx) \sim 5 \times 10^{-7}$. 
The term containing $\tan \varphi$ thus describes a distortion of the signal at the $\sim 10^{-6}$ level.

\subsection{Quantum limits for spin variances}
Different classical limits provide benchmarks for the radial and azimuthal components of a spin precessing in the $\Fy$--$\Fz$ plane. In general, these benchmarks describe the minimal noise of quantum states describing uncorrelated particles. For our system of $\NA$ spin-1 atoms, the lowest noise uncorrelated state is the \textit{coherent spin state} defined as a pure product state in which each atom is fully polarized in the same direction. If this direction is $\hat{y} \cos \theta - \hat{z} \sin \theta$, then the azimuthal component $F_\theta = -F_{y} \sin \theta + F_{z} \cos \theta$ has variance
\begin{equation}
\var(F_\psi)_{\rm SQL} = \frac{\langle F_\rho\rangle}{2}.
\end{equation}
Any state that surpasses this limit implies entanglement among the atoms, and/or entanglement of the internal components of the individual atoms\pciteSI{GuhnePR2009,SorensenPRL2001}

For the radial component $F_\rho = F_{y} \cos \theta + F_{z} \sin \theta$, the classical limit comes from the fact that accumulation of independent atoms into the ensemble is limited by  Poisson statistics, $\var(\NA) = \langle \NA \rangle$, so that for $F=1$, 
\begin{equation}
\var(F_\rho)_{\rm Poisson} = \langle \NA \rangle.
\end{equation}
Noise below this level can be produced by a strong interaction among the atoms during accumulation\cciteSI{SchlosserN2001,HofmannPRL2013,BeguinPRL2014,GajdaczARX2016} or as here by precise non-destructive measurement\pciteSI{StocktonThesis2007,TakanoPRL2009,AppelPNAS2009,SchleierSmithPRL2010,SewellPRL2012,BohnetNPhot2014,HostenN2016}

\subsection{Operator-level description of back-action evading measurement of two non-commuting spin observables}

We consider a spin variable ${\F}$, defined by commutation relations $[\Fx, \Fy] = i \Fz$ and cyclic permutations, precessing about the $\Fx$ axis and subjected to brief, non-destructive measurements of the $\Fz$ variable. 
We assume the precession during the measurement is negligible. 
In the measurement, the spin is coupled to the polarization of a probe pulse, described by the Stokes operators ${\bf S}$ with $[\Sx,\Sy] = i \Sz$ and cyclic permutations. 
The probe initial state is a coherent state polarized along $\Sx$, so that $\norm{\ave{\Sx}}=\NL/2$, $\ave{\Sy} = \ave{\Sz} = 0$, and $\var(\Sy) = \var(\Sz) = \frac{1}{2} \norm{\ave{\Sx}}$. 
The system and meter are coupled by the quantum non-demolition hamiltonian
\begin{equation}
H_{\rm eff} = g \Sz \Fz
\end{equation}
which acts for unit time. 
The transformation produced is 
\begin{eqnarray}
\Sx' & = & \Sx \cos g \Fz - \Sy \sin g \Fz \\
\label{eq:SyInOut}
\Sy' & = & \Sy \cos g \Fz + \Sx \sin g \Fz \\
\Sz' & = & \Sz \\
\Fx' & = & \Fx \cos g \Sz - \Fy \sin g \Sz \\
\Fy' & = & \Fy \cos g \Sz + \Fx \sin g \Sz \\
\Fz' & = & \Fz 
\end{eqnarray}
Where primes indicate the output variables.

We assume a spin state in the $\Fy$-$\Fz$ plane, i.e. with $\langle \Fx \rangle = 0$, and with zero initial cross-correlation, i.e. $\cov(\Fx, \Fy) = \cov(\Fx,\Fz) = 0$. 
Due to the zero mean of $\Sz$, which is also independent of ${\bf F}$, the transformation preserves these statistics in the primed variables, for example 
\begin{eqnarray}
\cov(\Fx', \Fy') 
& = & \cov(\Fx,\Fy) \langle \cos^2 g \Sz - \sin^2 g \Sz \rangle 
\nonumber \\ && 
+ [\var(\Fx) + \var(\Fy) ] \langle \cos g \Sz \sin g \Sz \rangle
\nonumber \\ & = & 0 
\end{eqnarray}
We can compute the statistics of the output variables using 
\begin{eqnarray}
\langle \cos g \Sz \rangle 
&= & \langle 1 - \frac{1}{2}g^2 \var(\Sz) + O(g)^4 \rangle 
\nonumber \\
&=& 1 - \frac{1}{4} g^2 |\langle \Sx \rangle| + O(g)^4 
\end{eqnarray}
and similar expansions for $\langle \cos^2 g \Sz \rangle$ and $\langle \sin^2 g \Sz \rangle$. 
The mean of $\Fy$ changes due to the back-action as
\begin{eqnarray}
\langle F'_y \rangle 
& = & \langle \Fy \rangle \langle \cos g \Sz \rangle + \langle \Fx \rangle \langle \sin g \Sz \rangle 
\nonumber \\ 
& = & \langle \Fy \rangle \langle \cos g \Sz \rangle
\nonumber \\ 
& = & \langle \Fy \rangle - \frac{1}{4} g^2 |\langle \Sx \rangle| \langle \Fy \rangle + O(g)^4
\end{eqnarray}
while the means of $\Fx$ and $\Fz$ are unchanged. 

The variance of $\Fx$ is coupled to the variance of $\Fy$, due to the rotation about $\Fz$ by a random angle $g \Sz$:
\begin{eqnarray}
\label{eq:vFxp}
\var(\Fx') 
& =& \langle (\Fx \cos g \Sz - \Fy \sin g \Sz)^2 \rangle - \langle \Fy \rangle^2 \langle \sin g \Sz \rangle^2 \nonumber \\ 
& = & \var(\Fx) \langle \cos^2 g \Sz \rangle+ \langle \Fy^2 \rangle \langle \sin^2 g \Sz \rangle
\nonumber \\ 
& = & \var(\Fx) + g^2 |\langle \Sx \rangle| \left[- \frac{1}{4} \var(\Fx) + \frac{1}{2}\langle \Fy^2 \rangle \right] 
\nonumber \\ 
& & + O(g)^4
\end{eqnarray}
and similarly
\begin{eqnarray}
\label{eq:vFyp}
\var(\Fy') 
& =& \langle (\Fy \cos g \Sz + \Fx \sin g \Sz)^2 \rangle - \langle \Fy \rangle^2 \langle \cos g \Sz \rangle^2 
\nonumber \\ 
& =& \langle \Fy^2 \cos^2 g \Sz \rangle + \langle \Fx^2 \sin^2 g \Sz \rangle - \langle \Fy \rangle^2 \langle \cos g \Sz \rangle^2 
\nonumber \\ 
& =& \langle \Fy^2 \rangle \langle \cos^2 g \Sz \rangle - \langle \Fy \rangle^2 \langle \cos g \Sz \rangle^2 \nonumber \\ & & + \langle \Fx^2\rangle \langle \sin^2 g \Sz \rangle 
\nonumber \\ 
& = & \var(\Fy) + g^2 |\langle \Sx \rangle| \left[- \frac{1}{4} \var(\Fy) + \frac{1}{2}\var (\Fx) \right] 
\nonumber \\ 
& & + O(g)^4
\end{eqnarray}
after noting that, to order $g^3$, $\langle \cos^2 g \Sz \rangle = \langle \cos g \Sz \rangle^2$.

After the coupling, a projective measurement of $\Sy'$ provides information about $\Fz$, with readout variance
\begin{eqnarray}
\var_{\rm RO}(\Fz) &\approx& \frac{ 1 }{2 g^2 |\langle \Sx \rangle| }.
\end{eqnarray}
The approximation comes from a linearization of Eq.~(\ref{eq:SyInOut}), which as discussed in Sec.~\ref{sec:FarRot} introduces an error at the $10^{-6}$ level, negligible in this scenario.

The resulting $\Fz$ variance, including both the prior and posterior information, is then \mciteSI{MadsenPRA2004,Colangelo2013a}
\begin{eqnarray}
\var(\Fz') 
& = & \frac{1}{\var^{-1}(\Fz) + \var_{\rm RO}^{-1}(\Fz) }
\nonumber \\ 
& = & \frac{\var(\Fz)}{1+ 2 g^2 |\langle \Sx \rangle| \var(\Fz)} 
\label{eq:vFzpth}
\end{eqnarray}
expanding in $g$ becomes
\begin{eqnarray}
\var(\Fz') & = & \var(\Fz) - 2 g^2 |\langle \Sx \rangle| \var^2(\Fz) + O(g)^4.
\label{eq:vFzp}
\end{eqnarray}

Collecting Eqs. (\ref{eq:vFxp}), (\ref{eq:vFyp}) and (\ref{eq:vFzp}), defining {$\Delta \langle F_\alpha\rangle \equiv \langle F'_\alpha\rangle - \langle F_\alpha\rangle $ and} $\Delta \var(F_\alpha) \equiv \var(F'_\alpha) - \var(F_\alpha)$, and dropping terms of order $O(g)^4$ we find
\begin{eqnarray}
\Delta \langle \Fy \rangle & = & -\frac{1}{2} g^2 | \langle \Sx \rangle | \langle \Fy \rangle \\
\Delta \var(\Fx)
& = & g^2 |\langle \Sx \rangle| \left[- \frac{1}{4} \var(\Fx) + \frac{1}{2}\langle \Fy^2 \rangle \right] \\
\Delta \var(\Fy)
& =&g^2 |\langle \Sx \rangle| \left[- \frac{1}{4} \var(\Fy) + \frac{1}{2}\var (\Fx) \right] \\
\label{eq:DvFz}
\Delta \var(\Fz)
& = & - 2 g^2 |\langle \Sx \rangle| \var^2(\Fz) 
\end{eqnarray}
Considering an initial coherent spin state and choosing $|\langle \Sx \rangle| = g^{-2} \NA^{-1}$, we note that $\Delta \var(\Fz)\sim \NA$, implying a reduction in the uncertainty of $\Fz$ comparable to its initial uncertainty. 
Due to the $\langle \Fy^2 \rangle$ term, the increase in $\var(\Fx)$ is $\sim \NA$, comparable to its initial value. 
The other changes are $\sim 1$, negligible relative to the initial values. In this way we see that uncertainty is moved from $\Fz$ to $\Fx$ with negligible effect on $\Fy$. 

Larmor precession then noiselessly rotates uncertainty from $\Fy$ into $\Fz$, uncertainty that is moved into $\Fx$ by the next measurement. 
This procedure reduces the uncertainty of both $\Fy$ and $\Fz$ with negligible influence from measurement back-action.

\subsection{Implementation in an atomic ensemble}
\subsubsection{Experimental set up}
The experimental set up is described in detail in references \citeSI{KubasikPRA2009,KoschorreckPRL2010a}.
The trap consists of a single beam laser at \unit{1064}{\nano\meter} with \unit{6.3}{\watt} of optical power, focused to a beam waist of \unit{26}{\micro\meter} using an \unit{80}{\milli\meter} lens.
The trap is loaded with laser-cooled atoms from a magneto optical trap (MOT). 
After sub-doppler cooling in the final stage of the loading sequence, the trapped atoms have a temperature \unit{\sim12}{\micro K}.
The resulting atomic ensemble has an approximately lorentzian distribution along the trap axis (which we label the $z$-axis) with a FWHM of $w=\unit{4}{\milli\meter}$, and a gaussian distribution in the radial direction with of $\omega=33\pm 3 \micro\meter$.

\subsubsection{State preparation}
\begin{figure}[t]
\includegraphics[width=\columnwidth]{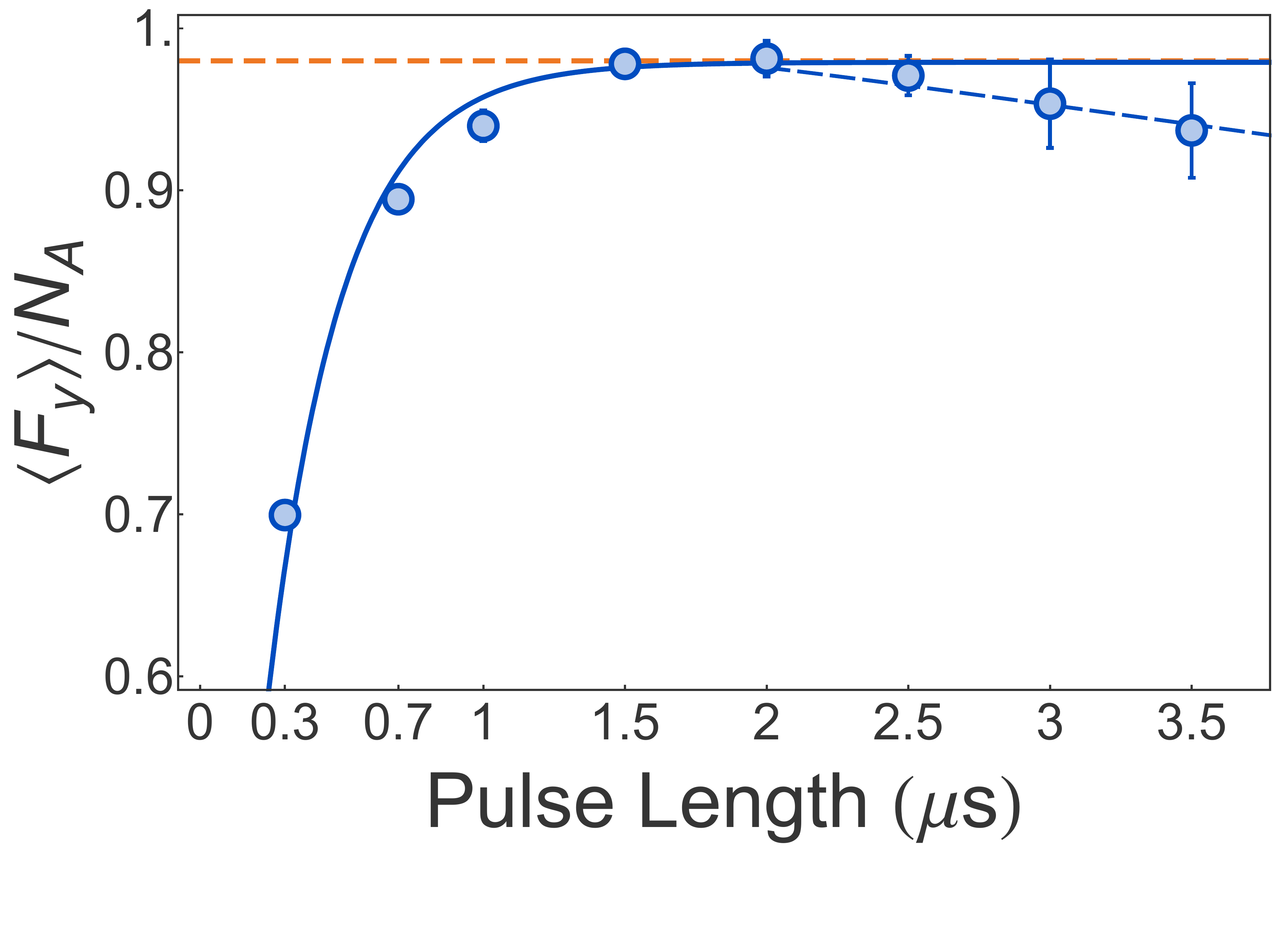}
\caption{
Optical pumping efficiency for the $\ave{\Fy}$ polarized state with a transversal field oriented in the $x$-direction. 
Data is fit with an exponential growing curve $\sim a(1- e^{-t/\tau})$ (solid line) and we obtain $a=0.979\pm 0.004$ and $\tau=0.26 \pm 0.02$. 
Orange dashed line: Optical pumping efficiency of $98\%$.
$\pm1\sigma$ statistical error bars are smaller than the points for most of the data.
\label{fig:OPFyBx}}
\end{figure}

The initial atomic state is prepared via optical pumping with circularly polarized light resonant with the $F=1\rightarrow F'=1$ transition propagating along the $y$-axis.
During the optical pumping stage the atoms are also illuminated with repumping light resonant with the $F=2\rightarrow F=2'$ transition using the six MOT beams, preventing accumulation of atoms in the $F=2$ hyperfine level, and a small magnetic field is applied along the $x$-axis, with $B_x=\unit{5.29}{\milli G}$, to coherently rotate the atomic spins in the $y$--$z$ plane.
We use a stroboscopic pumping strategy, chopping the optical pumping light into a series of $\tau_{\rm pump}=\unit{1.5}{\micro\second}$ duration pulses applied synchronously with the precessing atoms for total of \unit{200}{\micro\second}, to prepare the atoms in an $\Fy$-polarized state with high efficiency ($\sim98\%$), resulting in a input polarized atomic ensemble with $\ave{\Fy}\simeq\NA$ (see Fig.~\ref{fig:OPFyBx}).
The pulse duration $\tau_{\rm pump} \ll T_L$ is chosen to optimize the optical pumping efficiency.

\subsubsection{Probing}
We probe the atoms via off-resonant paramagnetic Faraday-rotation using $\tau=\unit{0.6}{\micro\second}$ duration pulses of linearly polarized light with a detuning of \unit{700}{\mega Hz} to the red of the \rb D$_2$ line. 
The probe pulses are $V$-polarized,
with on average $\NL=2.74 \times 10^6$ photons, and sent through the atomic cloud at \unit{3}{\micro\second} intervals.
Between the probe pulses, we send $H$-polarized compensation pulses with on average $\NLH= 1.49 \times  10^6$ photons through the atomic cloud.
As described in detail in references \citeSI{KoschorreckPRL2010b,SewellNP2013,Colangelo2013a},  the compensation pulses serve to cancel effects due to the tensor light shift, but do not otherwise contribute to the measurement.
During the probing sequence, a magnetic field along the $x$ direction drives a coherent rotation of the atoms in the $y-z$ plane with $T_L=\unit{38}{\micro\second}$ period. 
This ensures that the time taken to complete a single-pulse measurement is small compared to the Larmor precession period, i.e. $\tau \ll T_L$.

We correct for slow drifts in the polarimeter signal by subtracting a baseline $ {\varphi}_0 = \tfrac{1}{N} \sum_{k=1}^{N}\vph_{k}^{(i)}$ from each pulse, estimated by repeating the measurement without atoms in the trap.

\subsubsection{Statistics of probing inhomogeneously-coupled atoms}

We consider the statistics of Faraday rotation measurements on an ensemble of $\NA$ atoms, described by individual spin operators ${\bf f}_i$. 
To define the SQL, we consider an ensemble in a coherent spin state, with the individual spins are independent and fully polarized in the $\Fy$--$\Fz$ plane. 
We take $\NA$ to be Poisson-distributed. 
When the spatial structure of the probe beam is taken into account, the Faraday rotation is described by the input-output relation for the Stokes component $\Sy$
\begin{eqnarray}
\label{eq:FSQL}
\Sy\supout &=& \Sy\supin + \Sx\supin \sum_{i=1}^{\NA} g({\bf x}_i) \fz^{(i)} 
\end{eqnarray}
where $g({\bf x}_i)$ is the coupling strength for the $i$th atom, proportional to the intensity at the location ${\bf x}_i$ of the atom. $\Sy\supin$ is has zero mean and variance $|\langle \Sx\supin \rangle|/2$. 
We consider first the case in which the spin is orthogonal to the measured $\Fz$ direction, i.e. a measurement of the azimuthal component. Here the uncertainty in $g({\bf x}_i)$ and in $\NA$ make a negligible contribution, and the rotation angle $\varphi = \Sy\supout/\Sx\supin$ has the statistics
\begin{eqnarray}
\label{eq:mvphiOrthogonal}
\langle \varphi \rangle &=& \langle \fz \rangle \sum_{i=1}^{\NA} \left< g({\bf x}_i) \right>_{{\bf x}_i} 
\nonumber \\ 
& \equiv & \langle \NA \rangle \langle \fz \rangle \mu_1 \\
\label{eq:vvphiOrthogonal}
\var(\varphi) &=& \var(\varphi_0) + \var(\fz) \left< \sum_{i=1}^{\NA} g^2({\bf x}_i) \right>_{\NA,{\bf x}_i}
\nonumber \\ 
& \equiv & \var(\varphi_0) + \langle \NA \rangle \var( \fz) \mu_2
\end{eqnarray}
where $\varphi_0$ is the polarization angle of the input light, subject to shot-noise fluctuations and assumed independent of $\Fz$, and the angle brackets indicate an average over the number and positions of the atoms. 

Next we consider the case in which the spin is along the measured $\Fz$ direction, i.e., a measurement of the radial component. 
In this case, the uncertainty in $\fz$ is zero, and the variation in $g$ and in $\NA$ determines the measured variation
\begin{eqnarray}
\label{eq:mvphiParallel}
\langle \varphi \rangle &=& \langle \NA \rangle \langle \fz \rangle \mu_1 \\
\label{eq:vvphiParallel}
\var(\varphi) &=& \var(\varphi_0) + \langle \fz \rangle^2 \var\left( \sum_{i=1}^{\NA} g({\bf x}_i) \right)
\nonumber \\ 
& \equiv & \var(\varphi_0) + \langle \NA \rangle \langle \fz \rangle^2 v_2
\end{eqnarray}
We note that $v_2$ includes the variation of both the atom number and the coupling strength, and as such is lower-bounded by the Poisson statistics of $\NA$: $v_2 \ge \langle g^2({\bf x}) \rangle = \mu_2$.

For known $\langle \fz \rangle$ and $\var(\fz)$, measurements of $\langle \varphi \rangle $ and $ \var(\varphi)$ versus $\NA$ give the calibration factors $\mu_1$ and $\mu_2$ as described in Sections~\ref{sec:CalibrC} and~\ref{sec:CalibrQ}, respectively. 
To preserve the SQL $\var(\Fz) = \frac{1}{2} |\langle \Fy \rangle |$ and similar, in the analysis leading to Fig.~\ref{fig:PLPQS} we infer mean values as 
\begin{eqnarray}
\langle {F^{(a)}} \rangle & = & \frac{1}{\mu_1} \langle \varphi^{(a)} \rangle 
\end{eqnarray}
and covariances, including $\cov(A,A) = \var(A)$, as 
\begin{eqnarray}
\cov(F^{(a)},F^{(b)}) & = & \frac{1}{\mu_2} \cov( \varphi^{(a)}, \varphi^{(b)} ),
\end{eqnarray}
where $F^{(a,b)}$ and $\varphi^{(a,b)}$ are corresponding spin and angle variables. 
We note that because the contribution of $\var(\varphi_0)$ is not subtracted, this overestimates the spin variances.

\begin{figure}[t]
\includegraphics[width=\columnwidth]{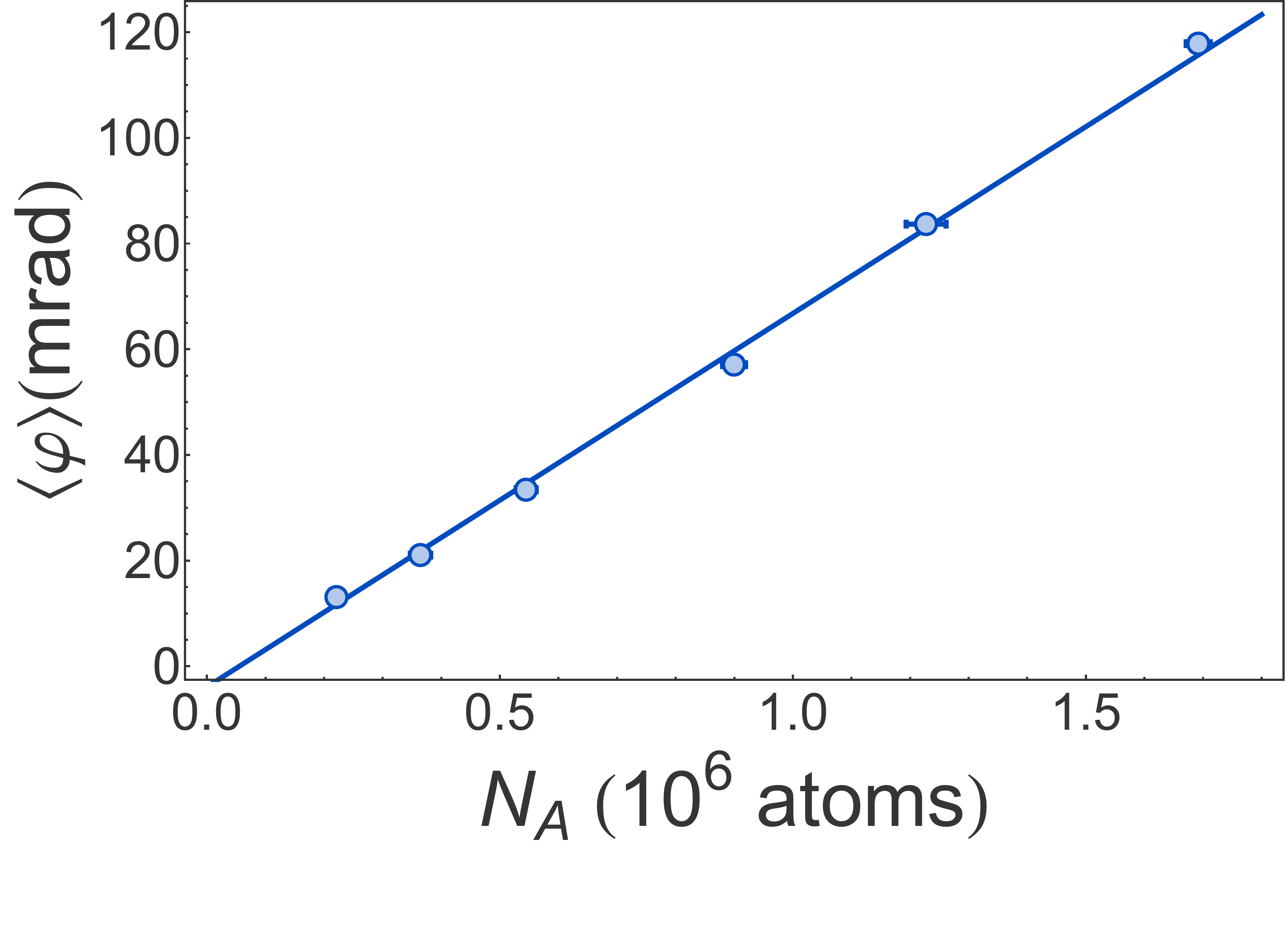}
\caption{Faraday rotation calibration using dispersive spin measurements by absorption imaging. Solid line, the fit curve $\vPh= a_0 + \mu_1 \NA$, with we obtain $\mu_1=(7.07\pm 0.04)\times  10^{-8}$ and $a_0=(3.9\pm 0.3)\times 10^{-3}$. Error bars indicate $\pm 1\sigma$ statistical errors.}
\label{fig:G1classical}
\end{figure}

\subsubsection{Measurement of calibration factor $\mu_1$}
\label{sec:CalibrC}
We calibrate the measured rotation angle $\varphi$ with a dispersive atom number measurements using absorption imaging, as shown in fig.~\ref{fig:G1classical}.
For the absorption imaging, atoms are transferred into the $f=2$ hyperfine ground state by a \unit{100}{\micro\second} pulse of laser light tuned to the $\gs(f=1)\rightarrow\es(f'=2)$ transition. 
The dipole trap is switched off to avoid spatially dependent light shifts.
An image is taken with a \unit{100}{\micro\second} pulse of circularly polarized light resonant to the $\gs(f=2)\rightarrow\es(f'=3)$ transition.
We calculate the resonant interaction cross-section and take into account the finite observable optical depth. 
The statistical error in the absorption imaging is $<3\%$, including imaging noise and shot-to-shot trap loading variation. 

\begin{figure}[t]
\includegraphics[width=\columnwidth]{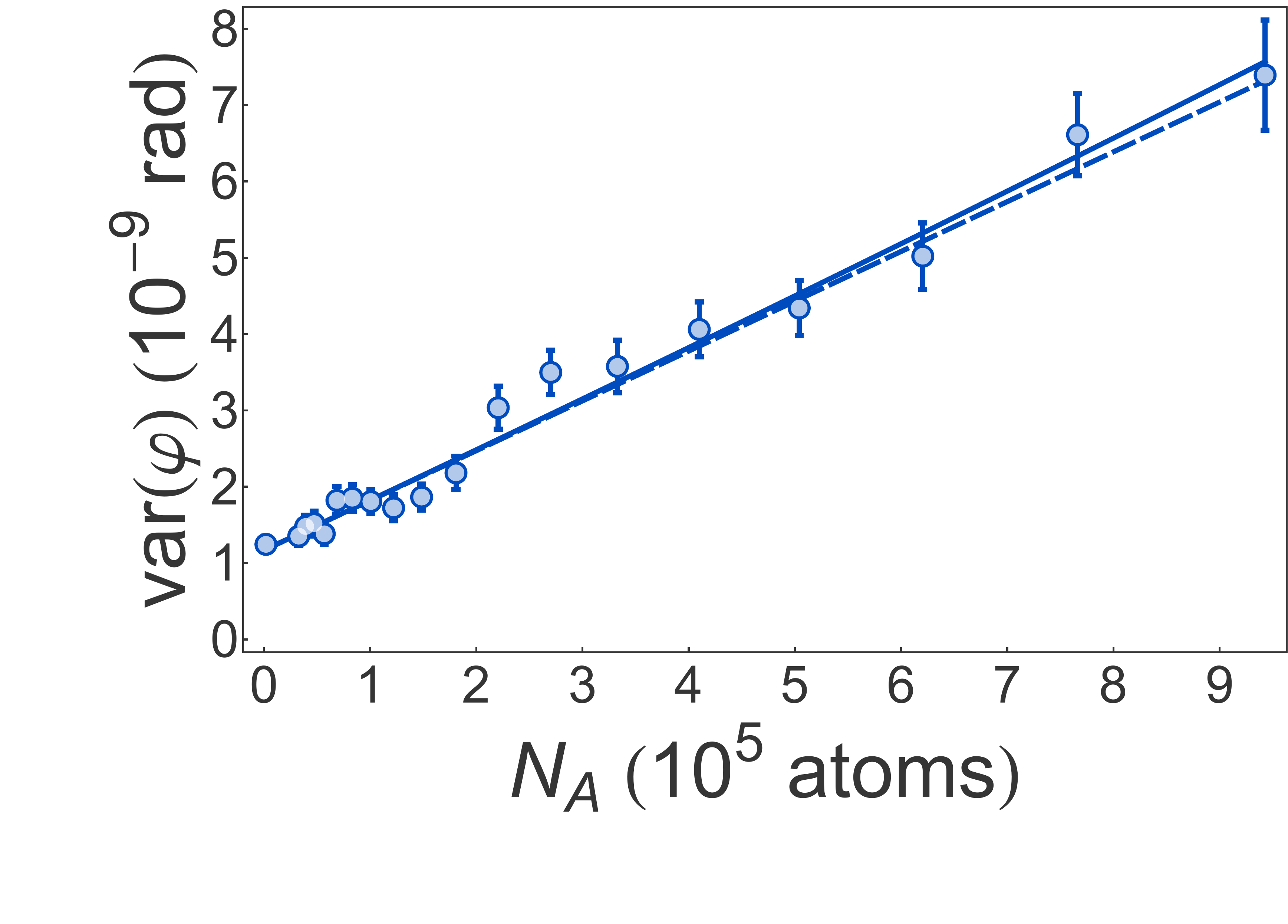} 
\caption{
Calibration of quantum noise limited Faraday rotation probing of atomic spins.
We plot the measured variance $\var(\vph)$ as a function of the number of atoms $\NA$ in an input $\Fy$ polarized state, i.e. $\ave{\Fy}=\NA$. 
Solid line: a fit using the polynomial $\var(\vph)=a_0 + a_1 \NA + a_2 \NA^2$ (solid line).
The linear term $a_1 = \alpha \mu_2 \NA/2$ corresponds to the atomic quantum noise from atoms in the input $\Fy$-polarized state. 
We estimate $a_0=(11.7\pm 0.7)\times 10^{-10}$,$a_1=(6.5 \pm 0.8)\times 10^{-15}$, and $a_2=(2.8\pm 12) \times 10^{-22}$, consistent with negligible technical noise in the atomic state preparation.
Dashed line: $\var(\vph) = a_0+a_1 \NA$.
Error bars indicate $\pm 1\sigma$ statistical errors. }
\label{fig:PNscaling}
\end{figure}

\subsubsection{Measurement of calibration factor $\mu_2$ }
\label{sec:CalibrQ}
To measure $\mu_2$ we prepare a $\Fy$-polarized state by optical pumping, and then probe stroboscopically with $\Npulse=36$ pulses of $\NL=3.15\times 10^{7}$ photons each in the presence of a B-field of $\approx \unit{71.5}{\milli G}$ along $y$, producing a Larmor precession of an angle $\pi$ during the \unit{10}{\micro\second} pulse repetition period. 
In this way, the measured variable is always $\pm \Fz$, evading back-action effects. 

If $\varphi_n$ is the measured Faraday rotation angle for pulse $n$, and $\varphi_0^{(n)}$ is the corresponding input angle, we can define the pulse-train-averaged rotation signal as 

\begin{eqnarray}
\vph &\equiv & \frac{1}{\Npulse}\sum_{n=1}^{\Npulse} (-1)^{n-1}\vph_{n} 
\end{eqnarray}

with variance
\begin{eqnarray}
\var(\vph) &\equiv & \var( \vph_0 ) + \mu_2 \sum_{n=1}^{\Npulse} \var(F_{z,n})
\end{eqnarray}
where $\vph_0 = \tfrac{1}{\Npulse}\sum_{n=1}^{\Npulse} \vph_0^{(n)}$, with zero mean and variance $\var(\vph_0) = (\Npulse \NL)^{-1}$, and ${F}_{z,n}$ is the value of $\Fz$ at the time of the $n$th probe pulse. 

During the measurement, off-resonant scattering of probe photons produces both a reduction in the number of probed atoms and introduces noise into ${\bf F}$. 
We note that this is a single-atom process that preserves the independence of the atomic spins. 
We compute the resulting evolution of the state using the covariance matrix methods reported in\cciteSI{Colangelo2013a} and specifically described for this case in Section~\ref{Sec:alpha}, giving
\begin{eqnarray}
\var\left( \frac{g}{\Npulse} \sum_{n=1}^{\Npulse} {F}_{z,n} \right) & = & \mu_2 \frac{1}{2} \alpha {\NA}
\end{eqnarray}
where $1/2 = \var(\fz) $ is the variance of the initial state, $\alpha = 0.86$ describes the net noise reduction due to scattering.

Including the readout noise $\var(\vph_0)$ and a generic technical noise $a_2 \NA^2$ in the preparation of the coherent spin state, we have the observable variance 
\begin{equation}
\var(\vph) =\var(\vph_0) + \mu_2 \frac{1}{2} \alpha {\NA} + a_2 \NA^2,
\label{varPHI}
\end{equation}
in which the $\NA$ scaling distinguishes the atomic quantum noise from other contributions. 
Experimental result shown in Fig.~\ref{fig:PNscaling} give $\mu_2=(1.5\pm 0.2)\times 10^{-14}$.

\subsubsection{Calculation of the noise contribution $\alpha$}
\label{Sec:alpha} 
As reported in ~\citeSI{Colangelo2013a},
the full system is described by a state vector $V=\{\Fz,\Sy^{(1)},\Sy^{(2)},\ldots,\Sy^{(N)}\}$ and covariance matrix $\Gamma = \ave{V\wedge V+(V\wedge V)^{T}}/2-\ave{V}\wedge\ave{V}$, where $\Sy^{n}$ is the measured photon imbalance after the $n$-th pulse.
The QND interaction leads to a transformation of the covariance matrix
\begin{equation}
\Gamma^{(n+1)} = {\bf M}^{(n)}\Gamma^{(n)}[{\bf M}^{(n)}]^{T}
\end{equation}
where ${\bf M}$ is equal to the identity matrix apart from the elements ${\bf M}_{1,1} = -1$ due to the precession by an angle $\pi$ about the magnetic field, and ${\bf M}_{n+1,1} = g S_{\rm x}$, where $S_{\rm x} = \NL/2$ and $\NL$ is the number of photons per pulse and $g$ is the coupling constant for uniform coupling.

Off-resonant scattering of photons introduces decoherence, noise and loss in the atomic state.
During the spin-noise measurement, a fraction $\xi=1-\exp(-\eta \NL)=0.01$ of atoms scatter a photon during a single probe pulse, where $\eta=3\times10^{-10}$ is the scattering rate per photon measured in an independent experiment, while a fraction $\chi=1-\xi$ remain in the coherent spin state. The scattered atoms are either lost from the $F=1$ manifold, or return to $F=1$ with probability $p=0.7$ and random polarization. 
This has the effect of losing atomic polarization at each measurement. 
We calculate the effective measured polarization in terms of the initial atom number.
We assume that the fraction $p$ of scattered atoms the return to $F=1$ have a random polarization and that the scattering rate $\eta$ is independent of the atomic state.

After each pulse, the atomic part of the covariance matrix transforms according to
\begin{equation}
\Gamma_{\rm at}^{(n+1)} = \chi \Gamma_{\rm at}^{(n)} + \frac{2}{3} p (1-\chi)\NA^{(n)}\mathbb{I}
\end{equation}
where $\mathbb{I}$ is the identity matrix.
This follows from Eq.(A.6) of \citeSI{Colangelo2013a} assuming $\Gamma_\Lambda = \NA \Gamma_\lambda$.
We note that we have
\begin{equation}
\NA^{(n+1)}=1-(1-\chi)(1-p)\NA^{(n)}=(\chi+p-\chi p)\NA^{(n)}
\end{equation}
which, assuming that $\NA^{(0)}=\NA$, gives
\begin{equation}
\NA^{(n)}=(\chi+p-\chi p)^n\NA
\end{equation}
Including these terms, we get a linear transformation of the covariance matrix after the $n$-th pulse
\begin{equation}
\Gamma^{(n+1)} = {\bf D}{\bf M}^{(n)}\Gamma^{(n)}[{\bf M}^{(n)}]^{T}{\bf D}^{T} + {\bf N^{(n)}}
\end{equation}
where ${\bf D}$ is a zero matrix apart from the element ${\bf D}_{1,1} =\sqrt{\chi} $, and ${\bf N^{(n)}}$ is the identity matrix apart from the element ${\bf N^{(n)}}_{1,1} =\frac{2}{3} p (1-\chi)(\chi+p-\chi p)^n \NA$.

We sum $N$ individual polarimeter signals $\Sy'^{(n)}$ to find the net Stokes operator $\Sy' \equiv \sum_{n=1}^{N} (-1)^{n-1} \Sy'^{(n)}$.
This has a variance
\begin{align}
	\var(\Sy^{'}) &= \sum_{n=1}^{N} \var(\Sy^{'(n)}) + 2 \sum_{n\neq m}^{N} \cov(\Sy^{'(n)},\Sy^{'(m)}) (-1)^{n-m} \nonumber \\
	&={\bf P}\Gamma^{(N)}{\bf P}
\end{align}
with the projector ${\bf P} = {\rm diag}(0,1,-1,1,-1,\ldots,-1)$.
When evaluated analytically using $\chi = 0.99$, this gives
\begin{equation}
\label{eq:varSyp}
\var(\Sy^{'}) =\var(S_{y,0}) + \beta g^2\frac{1}{2} \NA \NL^2
\end{equation}
where $\beta \approx 0.1081$.  Noting that $\var(\varphi) = \var(\Sy')/\Sx^2$, where $\Sx \equiv \sum_{n=1}^{N} \Sx^{(n)}$ is the total input Stokes operator, dividing Eq. (\ref{eq:varSyp}) by $\Sx^2=\NL^2/4$, and comparing against Eq. (\ref{varPHI}), we find that $\alpha = 8 \beta \approx 0.86 $.

\subsection{Data analysis}

\subsubsection{Conditional Covariance}

Estimating $\F$ for several values of $\tmeas$ gives a predictive trajectory and a confirming one. 
Estimations are repeated on 450 repetitions of the experiment to gather statistics.
Assuming gaussian statistics, to quantify the measurement uncertainty, we compute the conditional covariance matrix 
\begin{equation}
	\Gamma_{\F_2\mid \F_1}=\Gamma_{\F_2}-\Gamma_{\F_{2} \F_1} \Gamma_{\F_1}^{-1}\Gamma_{\F_1 \F_2}
	\label{eq:CondCov}
\end{equation} 
which quantifies the error in the best linear prediction of $\F_2$ based on $\F_1$ \pciteSI{BehboodPRL2014}
Here $\Gamma_{\bf v}$ indicates the covariance matrix for vector ${\bf v}$, and $\Gamma_{\bf uv}$ indicates the cross-covariance matrix for vectors ${\bf u}$ and ${\bf v}$.
Note that $A = \Gamma_{\F_{2} \F_1} \Gamma_{\F_1}^{-1}$ is identified as the matrix that minimizes the distance $\ave{ (\F_2 - A \F_1) D (\F_2 - A \F_1}$, where $D$ is a real symmetric matrix.
This suggests that we can visualize the difference between the best linear prediction of $\F$ using $\F_1$ and the confirming estimate $\F_2$ using the vector ${\bf \mathcal{F}}_k = \{ \mathcal{F}_{\rm y}, \mathcal{F}_{\rm z} \}_k = \F_2 - \Gamma_{\F_{2} \F_1} \Gamma_{\F_1}^{-1} \F_1$, where ${\bf F}_k^{(i)}=\{\Fy^{(i)},\Fz^{(i)}\}_k$.

\subsubsection{Fit Gain}
\begin{figure}
\includegraphics[width=\columnwidth]{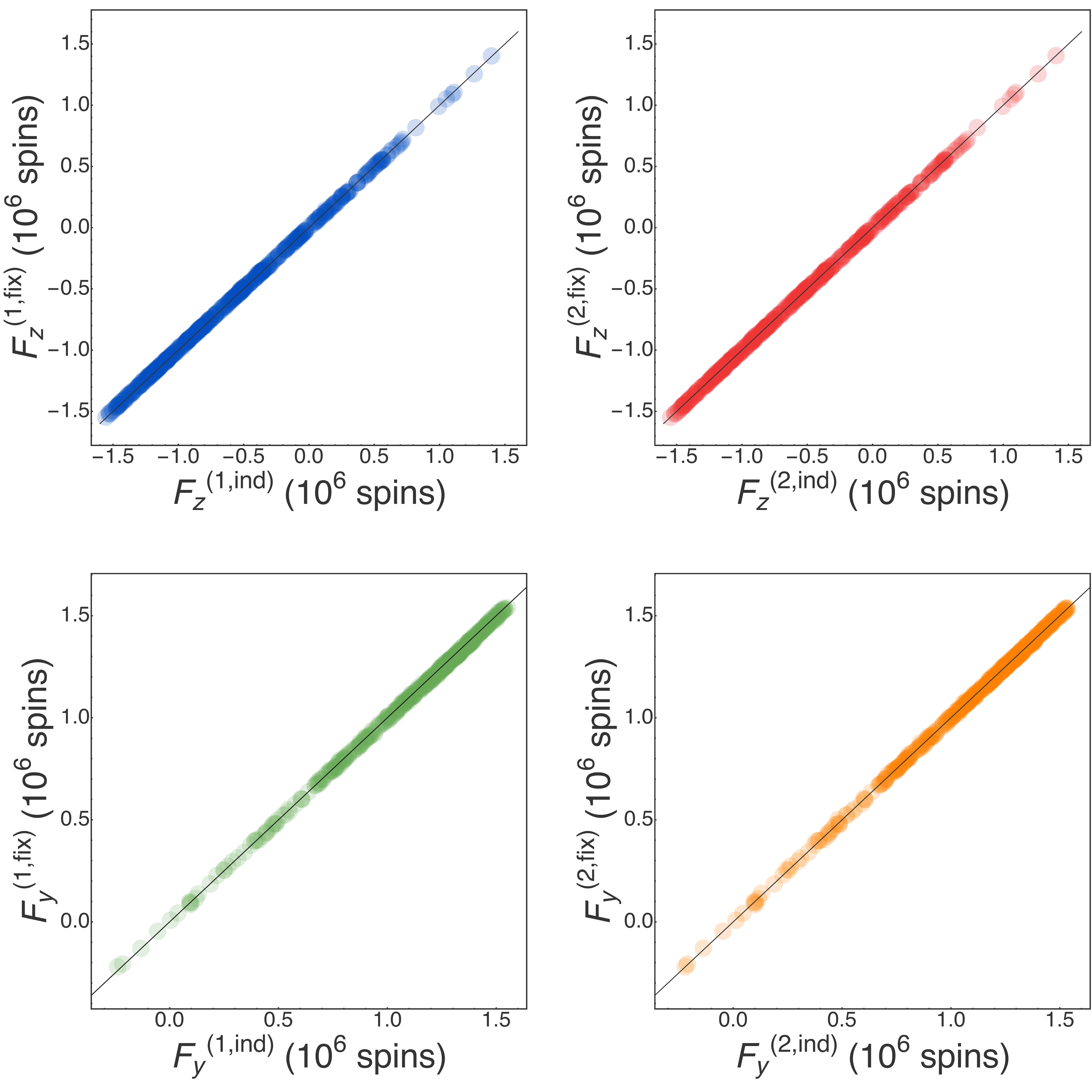}
\caption{
Comparison of the estimated $\Fz$ and $\Fy$ from a fit using Eq.~(\ref{eq:FIDForm}); 
first, with the classical parameters $g$, $\omegaL$, $T_2$ and $\offset$, fixed (labeled $\Fyz^{\rm (fix)}$) for measurements $1$ and $2$; 
and second, free to vary as independent parameters (labeled $\Fyz^{\rm (ind)}$).
In blue (green) $\Fz$ ($\Fy$) of the first measurement, in red (orange) $\Fz$ ($\Fy$) of the second measurement.
A linear fit $\gamma x+\delta$ to points of plots a-d gives $\gamma_{a}=0.9981(8)$, $\gamma_{b}=1.0026(8)$, $\gamma_{c}=0.9923(4)$, $\gamma_{d}=1.0007(5)$ and  $\delta_{a}=0.003(1)$, $\delta_{b}=0.0001(9)$, $\delta_{c}=0.0004(3)$, $\delta_{d}=-0.0023(3)$, where the subscripts refer to the values shown in plots a-d. 
A grey line $y=x$ is plotted on both the figures. 
\label{fig:FITgain}}
\end{figure}

Since the classical parameters $g$, $\omegaL$, $T_2$ and $\offset$ are fixed beforehand, the predictive and confirming fits are least-squares estimates obtained from disjoint data sets, optimized by minimizing the total variance ${\rm Tr}(\Gamma_{\F_2\mid \F_1})$.
We check that these fits give the correct gain by comparing the estimated $\F_{1,2}$ with the results of two independent fits using all free parameters in Eq.~(\ref{eq:FIDForm}).
Results, shown in figure~\ref{fig:FITgain}, indicate that the gains of the two fit procedures are equivalent at the $10^{-3}$ level.

\subsubsection{Weights}
As described in the main text, we follow a two-step fit procedure in our data analysis:
we first fit Eq.~\ref{eq:FIDForm} to the entire data set $\{\vph(\tpoint)\}$ to estimate the classical parameters $g$, $\omegaL$, $T_2$ and $\offset$ near the measurement time $\tmeas$;
then second, with the classical parameters fixed, we obtain a \textit{predictive} estimate $\F_1$ using measurements $\{ \varphi(\tpoint) \}_{\tmeas-\tdelta \le \tpoint < \tmeas}$; and a \textit{confirming} estimate $\F_2 $ using $\{\varphi(\tpoint)\}_{\tmeas < \tpoint \le \tmeas + \tdelta}$.

For the first fit to estimate the classical parameters, our data are weighted using an empirical function based on two observations: 
1) the polarimeter signal shows increased technical noise in the optical variable at larger imbalance, i.e. when measuring a large instantaneous spin-projection along the $z$-axis; 
and 2) points closer in time to $\tmeas$ should be given greater weight (minimizing errors introduced by small changes in $\omegaL$ and $T_2$ during the measurement).
This motivates using the weight function
\begin{equation}
	W(\vph(\tpoint)) \equiv \frac{g(\norm{\tpoint-\tmeas})}{h(\varphi(\tpoint))}.
\end{equation}
where $g(\norm{\tpoint-\tmeas})\equiv 1+A\exp\left(-w \left| \frac{t_k - t_e}{T_2}\right| \right) $ and $h(\varphi_k) = 1+r \, \varphi_k $.
We numerically optimize $W(\vph(\tpoint))$ varying the parameters $A$, $w$ and $r$ and minimizing the resulting ${\rm Tr}(\Gamma_{\F_2\mid \F_1}$) from the predictive and confirming fits.

For the predictive and confirming fits, which are linear in $\Fy$ and $\Fz$, all the points are weighted equally.

\subsubsection{Optimal measurement length}
\begin{figure}[h]
\includegraphics[width=\columnwidth]{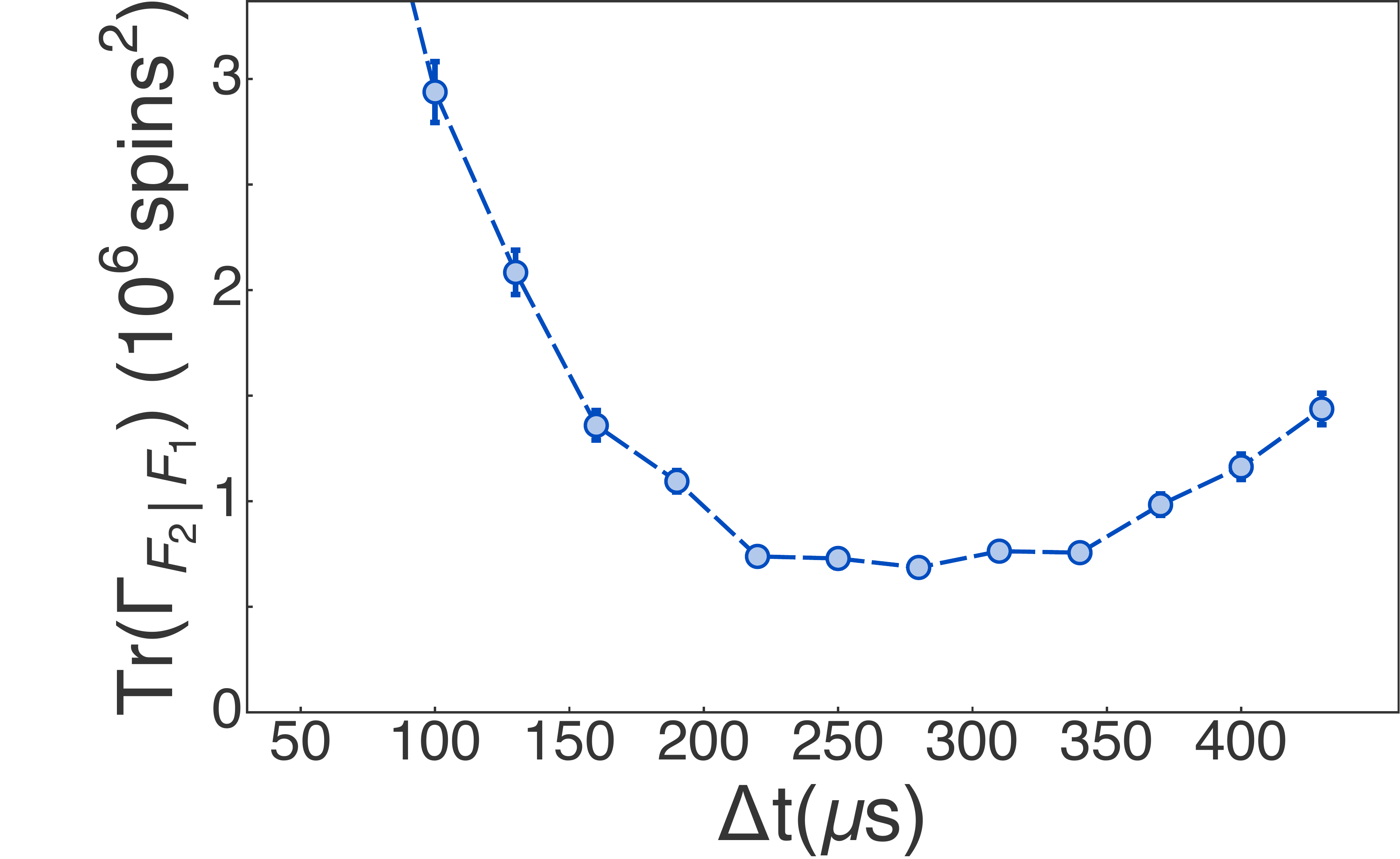}
\caption{Tracking precision as function of $\tdelta$. 
An optimum is found at $\tdelta=\unit{270}{\micro\second}$.
Error bars show $\pm 1 \sigma$ standard error. 
\label{fig:BestDt}}
\end{figure}

The optimal measurement length $\Delta t$ results from a trade off between the photon shot noise, the decoherences induced by the probing and the technical noise induced by the magnetic field. Longer measurements
reduces the photon shot noise, while increasing the atomic decoherences and making the model eq. \eqref{eq:FIDForm} less accurate. 
We empirically find the optimal $\Delta t$ by minimizing the total variance $\mathrm{Tr}(\Gamma_{\F_2\mid \F_1})$ for measurements with different length, as shown in Fig.~\ref{fig:BestDt}.

\end{document}